\documentclass[a4,12pt]{article}
\usepackage{graphicx}
\usepackage[2cell,arrow,curve,matrix,web]{xy}
\UseHalfTwocells
\UseTwocells
\usepackage{psfig}
\usepackage[dvips]{epsfig}
\begin{document}




\begin{titlepage}

\begin{flushright}

hep-th/0409016

\end{flushright}

\vspace{.5cm}

\begin{center}

{\Large \bf Holonomy Quantization of Moduli Spaces \\$\&$ \\
Grothendieck Groups}

\vspace{1cm}


I. Mitra{\footnote{indranil@theory.saha.ernet.in}$^a$\,}
\vspace{5mm}

{\em $^a$Department of Physics, New Alipore College\\
L Block, Kolkata 700 053, India}\\

\vspace{.5cm}
\end{center}

\vspace{.5cm}

\centerline{{\bf{Abstract}}}

\vspace{.5cm}

\begin{small}
Gelfand's \cite{Simmons.G1963}
 charecterization of a topological
space M by the duality relationship of M and $\mathcal{A} =
\mathcal{F}(M)$, the commutative algebra of functions on this
space has deep implications including the development of spectral calculas by Connes
\cite{Connes.A1994}.We investigate this scheme in this paper in the context of Monopole Moduli Space
$\mathcal{M}$
using Seiberg-Witten Equations \cite{Seiberg.N1994}.A observation has been made here that the methods
 of holonomy quantization using graphs can be construed to construct a C* algebra corresponding to the
 loop space of the Moduli. A map is thereby conjectured with the corresponding projectors of the algebra
  with the  moduli space.

\end{small}

\end{titlepage}

\newpage





\section{Introduction}
 Moduli spaces arising in string and field
theories, may be independent of the metric on the manifold.These
invariants prove very useful in manifold theory.In particular,
Seiberg and Witten \cite{Seiberg.N1994} gave a proof that the
instanton invariant of certain 4-manifolds  can be expressed in
terms of the Seiberg-Witten invariants.From the quantum field
theory point of view, the importance of Seiberg-Witten theorylies
in the concept of duality.In a modified version of Yang-Mills
theory, say N = 2 supersymmetric Yang-Mills theory, the quantum
field theory is described by a scale parameter and a complex
parameter u. In the a specific limit the theory is described by
an analytic function of u. Metric independence also implies
existence of a topological theories. Topological field theories
with extended
 topological symmetries have
appeared in various contexts for example in the discussion of S
duality in supersymmetric gauge theories \cite{R.Jackiw1984} as
world volume theories of Dirichlet p branes \cite{Polchinski1998}
in string theory and in a general discussion of balanced or
critical topological theories
\cite{R.Dijkgraff1991,Witten.E1988}. Unlike cohomological
topological field theories with an extended topological symmetry
topological theories (the most prominent example being Donaldson
Witten or topological Yang Mills theory
\cite{Witten.E1989,Witten.E1988a,S.K.Donaldson1996}) are
reasonably well understood. They typically capture the
deformation complex of some underlying moduli problem their
partition function is generically zero because of fermionic zero
modes while correlation functions correspond to intersection
pairings on the moduli space In \cite{Dijkgraaf1989a} there is an
investigation in some detail topological gauge theories of a
particular kind possessing a topological symmetry.These theories
typically capture the de Rham complex and Riemannian geometry of
some underlying moduli space and they have the characteristic
property of being critical i e of possessing a generically non
zero partition function equalling the Euler number of the moduli
space.  In some interesting works\cite{Vafa.C1995,T.Sasaki2000}
considered a particular topologically twisted version of N = d
super Yang Mills theory to perform a strong coupling test of S
duality of the underlying supersymmetric theory. Along similar
lines in \cite{1989}  a topological string theory calculating the
Euler number of the (Hurwitz) moduli space of branched covers was
constructed and shown to reproduce the large N expansion of two
dimensional Yang Mills theory.

 To make connection of holonomy quantization with
Grothendieck groups
 let us only mention
 in terms of topology,
that the orbifold fundamental groups of the moduli spaces of
curves and he explains that the first two levels principle is
essentially equivalent to the fact that the orbifold fundamental
group of any moduli space of dimension $> 2$
 is equal to its
fundamental group at infinity \cite{Patterson1979,Bers1979}. So
let us define the notion of fundamental group at infinity. If M
be a paracompact differentiable manifold and let us partially
order the compact submanifolds of M by inclusion. We need a base
point for our fundamental group and exploit the fact that a
fundamental group need not be based at a point but in fact at any
simply connected subset of the immersing manifold. Here a base
point at infinity, simply denoted by $x_{0}$, is given by an open
set $U \in M$ such that for any compact set $K\in M$, such that
$U/K$ is nonempty and simply connected. Let $p$ denote as usual
the topological fundamental group. There is a well known result
\cite{Lee1991}which states that quantum holonomy in a
three-dimensional general covariant non-Abelian gauge field
theory possesses topological information of the link on which the
holonomy operator is defined. The holonomy operator was shown to
be a central element of the gauge group so that, in a given
representation of the gauge group, it is a matrix that commutes
with the matrix representations of all other operators in the
group.  Quantum holonomy should therefore in general have more
information on the link invariant than the quantum Wilson loop
which, for the SU(2) Chern-Simons quantum field theory, was shown
by Witten  to yield the Jones polynomial \cite{Witten.E1989} .
Now it will be relevant from our point of view to consider some
typical properties of manifolds.We will denote by (T g,n) the
Teich\"{m}uller space of compact Riemann surfaces of type (g, n),
that is those which are obtained from surfaces of genus g by
marking n points. Let M g,n be the fine moduli space of surfaces
of type (g, n), which is obtained as the quotient of T g,n by the
(Teich\"{m}uller) modular group G g,n . More precisely G g,n acts
properly and discontinuously on T g,n with quotient M g,n . Since
the action is not free the latter space naturally inherits an
orbifold structure thus getting a singular manifold M g,n , which
is a coarse moduli space for the surfaces of type (g,
n)\cite{F.P.Gardiner2000}.

Now our motivation in this paper is twofold. We are concerned
here with the holonomy quantization of the monopole loop space
and prove that there exists a C* algebra corresponding to it. The
loop space is subsequently decomposed into automorphisms and a
toeplitz algebra. Construction of projectors of the algebra is
hence proposed and a physical interpretation is tried for. Our
main purpose in this paper is regards classification of moduli
spaces in context to the loop space algebra and K-theoretic
objects,K-theory in various forms has recently received much
attention in 10-dimensional superstring theory \cite{Witten1998}.
Twisted K-theory led to the discovery that it enters typically
into 3-dimensional topological field theories, in particular
Chern-Simons theory. Let G be a compact Lie group. There is a
theorem \cite{Freed1993,Freed1999} that for any compact group G,
a central extension of the free loop group is determined by the
level , which is a positive integer k. There is a finite set of
equivalence classes of positive energy representations of this
central extension, $\chi k(G)$ denote the free abelian group they
generate. One of the influences of 2-dimensional conformal field
theory on the theory of loop groups is the construction of an
algebra structure on $\chi k (G)$, the fusion product.  Since the
charges in TypeI String Theory have been shown to live in
$K_{0}$-theory, the 2-form gauge field is naturally interpreted
in differential $K_{0}$-theory. But here there are background
electric and magnetic currents which are present even in the
absence of D-branes. Their presence is most naturally explained
in a framework by the observation that the $K_{0}$~quadratic form
which defines the self-duality constraint is not symmetric about
the origin \cite{Moore1998}. For spacetime of the form $S = M^{d}
\times T^{n}$ where M is a d dimensional Minkowski spacetime and
T an $n$-dimensional manifold the $K_{0}$ constraint is no new
information if $n\le 7$.

In the context of algebraisation it should be noted that Reimann
moduli space has been extensively studied in recent years using
arc operads and algebras which has given rise to
Batalin-Vilkovisky and gravity algebras
\cite{R.M.Kauffman2003,A.Voronov2001}. So with all these
background we organize the paper as follows. In section
\ref{secgag} we review the basics of Wilson loop and Holonomy
quantization \cite{A.M.Polyakov1987,T.Busher1988}. In section
\ref{secsw} the Seiberg-Witten monopole equations are reviewed
and analyzed in the context of holonomy groups and the basic
geometrical structure of the moduli space is studied which will be
essentially important in proving the algebraic relations using
operator algebra's. In the last section \ref{secbasc} we prove the
C* algebras for the loop space for moduli space utilising the
basic insights from the previous analysis and thereby construct
the suitable projectors.

\section{Geometry of Gauge Fields  $\&$ Holonomy
Quantization:}\label{secgag}

\subsection{General Overview Of Wilson Loop:}

Fields are thought to be functions on a Principal bundle with
values in a vector space V,on which the gauge group G acts. Let
us have a gauge theory with gauge group G on a spacetime M. Let Y
denote the corresponding gauge field with values in the Lie
algebra of G. We also consider continuous curves in M whih are
piecewise differentiable , and call closed curves starting and
ending at point x a loops. The loops based at point x form a
group under product composition. Parallel transport around an
arbitrary loop in the base space leads to the concept of Holonomy
group $\textsl{Hol}(Y,C) \in G$ of the connection Y along C.
Holonomy is also very much related with curvature $\Omega$ of a
connection as has been established by\\
Let us now define the gauge functional Wilson loop to be

$ W_{\textit{G}}(C) Y = Tr\mid_{\textit{G}}(\textsl{Hol}(Y,C))$
where \textit{G} is a finite dimensional representation of G.

A physical way of thinking of Wilson loop operator is to identify
it as follows : If C is a loop in M then $W_{\textit{G}}(C)$ can
be considered as a function on $\Lambda/G$ ( where $\Lambda$ is
the space of gauge fields on M and G is the group of gauge
transformations on M) and as an operator in Hilbert
space.Consequently the Wilson loop can be conveniently
represented by a path integral, which will be of considerable use
in the techniques of holonomy Quantization.

Let C be a loop in M and let Y be a connection on the principle G
bundle P over M. Let \hspace{2cm}$f: S^{1}\longrightarrow C$ be a
parametrisation. Then we have\\
\begin{equation}\label{}
  W_{R}(C)(A) = \int_{\phi  \in  \Gamma({f^{*} P}/T)}e^{-I(\phi,A)}D\phi
\end{equation}
where $\phi$ is a section of $P/T$. {\footnote {\textit{G} is an
irreducible f dimensional representation of simle comact Lie
group G. By Borel Weil theory we can think of \textit{G} as
$H^{0}(G/T, {\textsl{L}}_{R})$ where ${\textsl{L}}_{R}$ is a
holomorphic line bundle.}}
\subsection{ Holonomy Quantization:}

Mandelstam\cite{Mandelstam.S1968} initiated holonmy approach for
the quantization of gauge theories. Consequently it was proposed
\cite{Mayer.M.E1977}to consider holonomy group as the object to
be quantized in gauge theory. In a nutshell the essential point
of holonomy quantization of gauge theories rests on the fact of
taking loops as the fundamental geometric elements in spacetime
\cite{Gervais.J1979,Nambu.Y1979,Makeenko.Yu.M1981}. Interestingly
the situation can be compared to the closed string theory.

In the following we describe in brief the details of Mandelstam
approach which we will extend to our construction in section 5.
When connection is given on the manifold,parallel transport of a
vector along a curve on a manifold is defined and so is the
parallel transport of the frame along the curve on the manifold.
Now choose a closed curve through a point p. Parallel transport
of the given frame along the closed curve starting at p and
returning to p will lead to a new frame at p. this new frame is a
linear transformation of the old frame so that the parallel
transport of a frame along a closed curve is an automorphism of
the tangent space $T_{p}M$. For different closed curves through p
we have different automorphisms of the tangent space $T_{p}M$.
The set of automorphisms of $T_{p}M$ constitutes the Holonomy
group with respect to the reference point p. For a connected
manifold, the holonomy group is independent of the choice of the
distinguished point chosen and is determined by the topological
properies of the manifold. In the abelian case{\footnote {The non
abelian case was treated by
\cite{Biyalynicki-Birula.I1963}}}defined by a U(1) principal
bundle on Minkowski space the holonomy groupoid consists of all
parallel transports along the paths where the end of the first
coincides with the begining of the second to be composed, leading
to a new element ; the inverse is the parallel transport along the
opposite path. Now here it is convenient to think of spacelike
onepoint compactification of M which is done by assuming the
existence of a single point at spatial infinity.\\
The following two theorems classify the manifolds geometrically
on the basis of holonomy:

\textbf{Theorem }(De Rahm)\textbf{:}\\
Let M be a compact simply connected Reimannian manifold. There
exists a cannonical decomposition M$
\stackrel{\sim}{\longrightarrow} \prod_{i}M_{i}$ where each
$M_{i}$ is an irreducible manifold. Let $p = (p_{i})$ be a point
of of M and let $H_{i}\subset O(T_{p}(M_{i})$ be the holonomy
group of $M_{i}$ at $p_{i}$ then the holonomy group of M at p is
the product $\prod H_{i}$ acting on $T_{p}(M) = \prod
T_{p_{i}}(M_{i})$ by the product representation.

We are thereby reduced to irreducible (compact, simple connected)
Reimannian manifolds,symmetric spaces of the form G/H where G is
a compact Lie Group $\&$ H is the neutral component of the fixed
locus of an involution of G.

\textbf{Theorem }(Berger)\textbf{:}

Let M be an irreducible (simply connected) Reimannian manifold,
which is not isomorphic to a symmetric space. Then the holonomy
group of M belongs to the following types:
\begin{center}
\begin{tabular}{|c|c|c|}
  H & dim(M) & metric \\
  SO(n) & n & generic \\
  U(m) & 2m & Kahler \\
  SU(m) $m\geq 3$ & 2m & Calabi Yau \\
\end{tabular}
\end{center}
After this brief digression, we can now state thereby that the
fields and Components of the curvature along holonomy transforms
along a spacelike path of reference fields defined at the special
point chosen. So if we start with conventional scalar
electrodynamics described by scalar field $\phi (x,t)$ satisfying
\begin{center}
\begin{eqnarray}\label{}
[(d_{+} - ieA)^{\star}(d_{+} - ieA) - m^{2}]\phi_{+} = 0\\
\nonumber F = dA ; dF = 0\\ \nonumber
d \ast dA + \ast J = 0
\end{eqnarray}
\end{center}
In terms of the special point chosen say $x_{0}$ and a particular
spacelike path P joining $x_{0}$ $\&$ x the holonomy transformed
fields are obtained by the holonomy operator.
\begin{center}
\begin{equation}
\Phi(x,P) = \Phi^{\dagger}(x_{0})W_{R}(C)(A) = {\Huge[\int
e^{-I(\Phi , A)}D\Phi ]\Huge} \Phi(x)
\end{equation}
\end{center}
The response of the functionals $\Phi(x,P)$ to a variation of
path P consisting of attaching an infitisimal loop spanning the
area vector $\sigma$ attached to the path at the point z,is given
by $\delta_{\sigma}\Phi(x,P) = -i W_{R}(C) [F] \Phi(x,P)$

The integrability conditions can be easily established since the
F's commute along different points of a spacelike path.(in the
nonabelian case there is an intrinsic noncommutativity)
{\footnote{The nonabelian case was treated by
\cite{Biyalynicki-Birula.I1963}}}

One can also define the directional derivatives
$\partial,\bar{\partial}$
\begin{center}
$ \partial \Phi (x,P) = W_{R}(C)(A)D\Phi \hspace{0.3cm};
\hspace{1cm}[\partial,\bar{\partial}]\Phi(x,P) = i e
\hspace{0.25cm}dA \Phi $
\end{center}
and immediately derive the equations of motion of fields $\Phi,
\Phi^{+}$ which look like free Klein Gordon equation with
derivatives being replaced by the noncommuting derivatives
$\partial$. Mandelstam \cite{Mandelstam.S1968}has shown that one
can get all the results of the Coulomb gauge formulation from
this approach. From these equations Mandelstam has succeeded in
deriving sets of equations for path dependent Green functions for
the model, which then can be expressed in terms of auxiliary path
independent quantities , i.e. the gauge dependent components of
the connection and curvature lead to same perturbation theory as
Feynman De-Witt approach.

Now the basic question whch we may ask at this stage is that what
has holonomy quantization has to do with our studies on moduli
spaces. since we are mainly interested in algenraic decomposition
of moduli spaces, the basic motivation stems from the fact that
we may associate a Wilson loop like holonomy operator for a path
or loop. $$W(P) = e^{\int_{P}A} = \lim \prod_{p} e^{A(x,
\triangle x)} = \lim \prod_{p} U(x, \triangle x)  \forall U\in
Hol(F,C)$$
The holonomy groupoid can be defined by implementing Wilson loop
or path integral or the equivalent operator associated with it,
which takes the form
\begin{equation}\label{a}
 U(P) \Phi(s) U^{-1}(P) =
 \Phi(W(P) s)
 \end{equation}
 where $s: M \rightarrow E$ is a test section of the associated
 vector bundle E, $\Phi$ is the operator valued field distribution
 on s. W(P) is the holonomy along a path P acting on the section
 s.
 It seems that (\ref{a}) may be impossible to implement via a
 unitary operator functionals on the path where thereby one has to
 consider general mappings of operators which give a
 representation of the holonomy groupoid by the automorphisms of
 the operator algebra. This observation as we shall see will be
 the crux of the whole story.

 \section{ Moduli Space Structure, Seiberg-Witten(SW) Equations $\&$
 Vacuum from Gauge Configurations}\label{secsw}
This section is mainly a review on monopoles and corresponding
 moduli spaces\cite{A.Sen1994,Prasad.M.K1975,M.F.Atiyah1988}.

\subsection{BPS $\&$ SW Equations}
 Yang-Mills(YM) potential A(x) corresponds
 to the connection on a principle bundle, the matter field
 $\phi(x)$ corresponds to the section of the associated bundle. If
 we consider an SU(2) gauge theory with spontaneous broken
 symmetry i.e. coupled to Higgs field with a potential allowing
 for non zero vacuum expectation values. The Lagrangian is
 $$\mathcal{L} = -\frac{1}{4}F^{(a)}_{\mu\nu}F^{\mu\nu}_{(a)} +
 \frac{1}{2}D_{\mu}\phi^{a}D^{\mu}\phi_{a} -
 \lambda^{2}{(\phi^{a}\phi_{a} - c^{2})}^{2}$$
 By rescaling in the Coulomb gauge $A_{0}$ the static energy
 density $\mathcal{E}(\partial_{0}\mathcal{E} = 0)$ can be written
 as
$$ \mathcal{E} = \mp  \int d^{3}x
\partial_{i}(H^{i}_{a}\phi^{a}) +
 \int d^{3}x  [ \frac{1}{2}{\| H^{i}_{a} \pm D_{i}\phi^{a}\|}^{2} +
 \lambda^{2}{(\phi^{a}\phi_{a} - 1)}^{2}]$$
 The first term on the RHS is gauge invariant which is
 propositional to an integer self induced conserved charge n,
 called the topological charge. By the +ve definite property of
 the last two terms we obtain $ \mathcal{E} \geq nQ_{m}$
 saturation of this inequality requires $ \lambda = 0$ which is
 the famous BPS limit which is equivalent to introducing a
 massless Higgs particle with a nonzero vacuum expectation value
 $\phi^{2}\rightarrow 1 $ The perturbative states consist of a
 massles photon, a massless neutral scalar and $W^{\pm}$ bosons
 with electric charge $Q_{e} = \pm e$
\noindent
 The construction of the static monopole solutions requires
 imposing the Gauss's law, the $A_{0}$ equation of motion as a
 constraint on the physical fields
$$D_{i}A_{i} + e[\phi ,\dot{\phi}] = 0$$ The case in hand turns
out to be
$$\mathcal{L} =
-\frac{1}{4}F^{(a)}_{\mu\nu}F^{\mu\nu}_{(a)} +
 \frac{1}{2}D_{\mu}\phi^{a}D^{\mu}\phi_{a}$$
$$\mathcal{E} = n +
 \int d^{3}x  [ \frac{1}{2}{\| H^{i}_{a} \pm
 D_{i}\phi^{a}\|}^{2}$$
So the static energy is minimised when the bound is saturated
which implies $$ H^{a}_{i} = \pm D_{i}\phi^{a}$$ called Bogomolny
equations. The +ve sign corresponds to monopoles and the -ve sign
is the counterpart for antimonoples.

\noindent The previous analysis is based on the study of
irreducible non abelian gauge theories with spontaneous broken
symmetry and Prasad Sommerfield limit. Now look at its
supersymmetric extensions.

\noindent Let (M,G) be a compact oriented 4 dimensional
Reimannian manifold with local spin structure $$ S = S^{+} +
S^{-}$$ Let $\Gamma(S^{\pm})$ denote the sections of bundles
$S^{\pm}$. The Dirac operator $D: \Gamma(S^{+})\rightarrow
\Gamma(S^{-})$ is given by $D = \sum e_{i}\nabla_{e_{i}}$ where
$\{e_{i}\}$ are the basis of Clifford algebra satisfying $$
e_{i}\cdot e_{j} + e_{j}\cdot e_{i} = -2 \delta_{ij}$$ The
product between $ e_{i}$ and $\nabla_{e_{i}}$ is Clifford product
and $\nabla_{e_{i}}$ denotes covariant derivative with Levi-Civita
connection on (M, g)
\begin{eqnarray}
D^{+}D\psi &=& \sum e_{i}. e_{j} (\nabla_{e_{i}}\nabla_{e_{j}} -
\nabla_{\nabla_{e_{i}e_{j}}})\psi \\ \nonumber &=& -\triangle \psi
+ \sum_{i<j} e_{i}. e_{j} [(\nabla_{e_{i}}\nabla_{e_{j}} -
\nabla_{\nabla_{e_{i}e_{j}}})(\nabla_{e_{j}}\nabla_{e_{i}} -
\nabla_{\nabla_{e_{i}e_{j}}})]\psi \\ \nonumber
&=& -\triangle
\psi + \frac{\mathcal R}{4}\psi
\end{eqnarray}
where $\mathcal{R}$ is scalar curvature of (M, g). The Spin field
may not be defined globally. A tangent bundle T(M) admits a spin
structure if there exists an open covering $\{U_{\alpha}\}$ and
the\\ $$h^{\alpha\beta} : U_{\alpha}\cap U_{\beta}\rightarrow
Spin(n)$$ can be chosen such that they satisfy the cocycle
condition $ h^{\alpha\beta}h^{\beta\gamma}h^{\gamma\alpha} = 1$
\noindent A manifold M is called a spin manifold if its tangent
bundle admits a spin structure. For spinors to be well defined on
T(M) i.e. spin structure exists iff its second Steifel Whiteney
class $\omega_{2}(T(M)) \in H^{2}(M, Z_{2})$ vanishes. If
$\omega_{2}(T(M)) \neq 0$ there is a sign ambiguity when spinors
are parallel transported around a closed path on M, which means
spinors are not well defined.

\noindent Now we introduce a complex line bundle L. The
classification theorem which states that there is a natural
one-one correspondence between the set of equivalence classes of
principal fiber bundles with given base M and structure group G
and the set $[M, B_{G}]$ of homotopy classes of mappings from M
to $B_{G}$ of a universal bundle ${\mathcal{E}}_{P}(G)$

\noindent So eventually we are given a k dimensional vector
bundle E over M and a map \\
$ f : M \rightarrow G^{r}_{r}(\infty,
k)${\footnote{$G_{r}(n, k)$ is the manifold whose points are
oriented k dimensional planes in $R^{n}$ passing through the
origin.}}, the pullback bundle $f^{*}L(\infty, k, C) = E$ and
$f^{*}C_{i}(L(\infty, k, C)) = C_{i}(E) = H^{k}(M, \prod_{k-1}G)$
is determined uniquely upto homotopy. So the Cohomology classes
are all well defined and depend on the bundle E only.

\noindent A general D=4 SUSY gauge theory with $N \leq 4$
supersymmetries  consists of the following field content: a gauge
field $A_{\mu}$ together with N majorana fermions
$\lambda^{I}_{a}$ $(I = 1,\cdots N)$, and $\frac{1}{2}N(N - 1))$
real scalar fields conveniently described by $N \times N$
antisymmetric matrix, $\phi^{IJ} = - \phi^{JI}$. All these fields
take action in the adjoint representation of the gauge group. The
simplest SUSY action is of the form \ $$ S = \int Tr ( F \wedge
\ast F + i \bar{\lambda}_{I} \phi \lambda^{I} + D
\phi_{IJ}\wedge\ast D \phi^{IJ} + \sum_{IK} {|\phi^{IJ},
\phi_{JK}|}^2
 + \lambda_{I}|\phi^{IJ}\lambda_{J}|)$$
This action has the symmetry group $Spin(4)\times U(2)_{R}$ where
$Spin(4)$ is the double cover of the Lorentz Group $SO(4)$. The
twisted SUSY algebra $\{Q, G_{\mu}\} = P_{\mu}$ gives an one-one
correspondence of the theory with topological field theory.

Now if we specialize for the case of obtaining monopole equations
we introduce a line bundle L on M satisfying $ c_{1}(L) =
\omega_{2}(M) (mod 2)$. A spin$^{C}$ structure is then a lift of
the fiber product of the $SO(4)$ bundle of orthonormal frames of
(M, g) with the U(1) bundle defined by L to a spin$^{C}$ bundle
according to the short exact sequence.\\
$$ 0\rightarrow Z_{2}\rightarrow Spin^{C}(4)\rightarrow SO(4)
\times U(1)\rightarrow 1$$ So we can be assured of a globally
defined Spin field $\Psi$ which is the section of the complex
vector bundle $W^{+}$\\
$$W^{+} = S^{+} \oplus L^{1/2} , L^{1/2} \otimes L^{1/2} = L$$
Let $ D_{A} : \Gamma(S_{L}^{+}) \rightarrow \Gamma(S_{L}^{-})$
where $ D_{A} = \sum e_{i}\nabla^{A}_{e_{i}}$ denotes the Dirac
operator where\\
$$\nabla^{A}_{e_{i}}(S^{+} \otimes L^{1/2} = \nabla^{LC}_{e_{i}}
\otimes L^{1/2} + S \otimes \nabla^{A}_{e_{i}} L^{1/2}$$ Here
$\nabla^{LC}_{e_{i}}$ denotes covariant derivative with Levi
Civita
connection of (M, g) and\\
$$\nabla^{A}_{e_{i}}l = A_{i}l ,
\hspace{1cm}\nabla^{A}_{e_{i}}l^{1/2} =
\frac{A_{i}}{2}l^{1/2}\hspace{1cm} l^{1/2} \in L^{1/2}$$ where A
is a U(1) valued connection 1 form on complex line bundle L, and
$L^{1/2}$ is another complex line bundle on M with $ L^{1/2}
\times L^{1/2} = L$. So the SW are eqns for a pair (A, $\Psi$)
consisting of a unitary connection on L and smooth section of
$S^{+}_{L}$.

The curvature $F_{A} \in \Omega^{2} = \Omega^{+} \oplus
\Omega^{-}$ can be decomposed into self dual and anti self dual
parts\\
$$F_{A} = F_{A}^{+} + F_{A}^{-}$$ One of the basic ingredients
which makes SW equations possible is the identification between
the space of self dual 2 forms and the skew hermitian
automorphisms of the +ve spin representation.

The spin field $\psi \in \Gamma(W^{+})$ is coupled to the $U(1)$
gauge field and the SW functional is constructed
\begin{equation}
SW(A, \psi) = \int_{M}\big( \|D_{A}\psi\|^{2} +\|F_{A}^{+}\|^{2}
+ \frac{\mathcal R}{4}\|\psi\|^{2} +
\frac{1}{8}\|\psi\|^{4}\big)d^{4}x
\end{equation}
$\|\psi\|^{2} = <\psi, \psi>$ is the invariant Hermitian product
on $\Gamma(W^{+})$. The Euler-Lagrange equations are
\begin{eqnarray}
&&\triangle_{A}\psi + \frac{\mathcal R}{4}\psi +
\frac{1}{4}\|\psi\|^{4}\psi = 0\nonumber\\
&& \delta F_{A}^{+} + \frac{1}{2}Im <\nabla_{i}\psi, \psi>e_{i} =
0
\end{eqnarray}
This is of the form of a nonlinear Schrodinger equation coupled
with a U(1) gauge potential. The absolute minima of SW functional
can be obtained which gives rise to the famous Seiberg Witten
Monopole equations which we term as moduli space charecterization
(MSC) equations.
\begin{eqnarray}\label{mono}
D_{A}\psi = 0 \quad;\quad\quad F_{A} = \frac{i}{4}<e_{i}e_{j}\psi,
\psi>e_{i}\wedge e_{j}
\end{eqnarray}
As we will note at this point that we propose that the alice
string behaviour on the monople moduli space requires unitary
operator implementation whose functional counterparts will
evidently satisfy (\ref{mono}). So in order to classify a moduli
space algebraiclly the first step is to construct the MSC
equations corresponding to the space. The operator implementation
will be done on the corresponding functionals obtained.

\subsection{ Geometry of Moduli Space \&\ Spaces of Loops in
$\mathcal M_{k}$}\label{geomod}
 The moduli space of gauge
inequivalent solutions of Bogumol'nyi equations are denoted by
$\mathcal M_{k}$. The configuration space of fields is given by
$\mathcal C = \mathcal{A}/\mathcal{G}$ where$ \mathcal A = \{
A_{i}(x), \phi_{x}\}$ is the space of finite energy field
configurations and $\mathcal G $ is the group of gauge
transformations that go to identity at spatial infinity.So as we
have seen earlier this point may be identified with the
distinguished point for the application of the principles of
holonomy quantization. The quantum moduli space which has been
considered by Seiberg \&\ Witten for massless fields whose
correlation functions do not decay exponentially with distance.
The moduli space of Higgs vacua is obtained by the relation
$<\phi_{a}\phi_{a}> = v^{2}$ and is thus a two sphere. If
$v^{2}\neq 0$ then SU(2) is spontaneously broken to U(1) in the
form $$ A = Tr(\hat{\phi} A)\quad;\quad\quad \hat{\phi} =
\frac{\phi}{\mid \phi\mid}$$ It is obvious from above that it
does not make sense at zeroes of $\phi$. Around such a zero the
unit vector field the unit field $\hat{\phi}$ defines a
nontrivial element of $\pi_{2}(S^{2})$ with winding number $w$,
which represents an obstruction to deform zero away. These
configurations are charecterised by the property that for a two
sphere in $R^{3}$ such a zero $ \int_{S^{2}} dA = 2\pi w$. So the
singularity carries possible magnetic charges.\\
Therefore the SW monopole equations give rise to the study of
moduli spaces. Following the analysis of Seiberg-Witten the
quantum moduli space is given by the ansatz $\mathcal M_{k} =
H/\Gamma(2)$ where $$\Gamma(2) = \{\gamma \in SL(2, Z)\} $$ which
is the moduli space of elliptic curves. In fact $\mathcal M_{k}$
parametrises elliptic curves $\sum $ of the particular form $
y^{2} = (x^{2}-1)(x-u)$.\\
Now an analysis by \cite{T.Imbo2004} has made an interesting
observation. As it is already known that if $\pi_{1}(V)$ is
nontrivial where V is the vacuum manifold and it is equivalent to
$G/H$ of a spontaneously broken gauge symmetry, the resulting
defects are strings. Such strings carry a topologically stable
flux which compensates the gradient of Higgs field at spatial
infinity. It has been thereby shown that with a suitable
$U(\phi)$, a Wilson line associated with the strings nonabelian
flux we construct the transformation of vacuum condensate as $$ <
\phi(\varphi)> = U( \varphi)<\phi>_{0}$$ The Alice Strings are
constructed by the states which transform as
\begin{equation}\label{alice}
 H(\varphi) = U( \varphi) H_{0} U^{-1}(\varphi)
\end{equation}
  It has been
shown in \cite{T.Imbo2003} that twisted Alice loops do give rise
to the existence of global monopoles.

So our studies regarding the moduli spaces and the connection of
Alice loops with monopoles give us a clue regarding the
construction of holonomy quantization or their extensions graphs
as we will see in section(\ref{secbasc}).
\section{The Basic Construction}\label{secbasc}
Our studies in Section 2 and 3 have led us to believe that the
moduli space geometry may be related to holonomy of paths in the
moduli space. As we will show now this idea can take us furthur.
In our quest for algebraization of geometry we thereby propose a
general model to probe the geometry of moduli spaces as we
understand today by constructing an operator algebra from the
holonomy of graphs, an extension of paths and loops. As a
specific example we will construct a holonomy groupoid for the
monopole moduli space and will thereby show that the holonomy
groupoid induces a C$^*$ operator algebra from which we have
constructed the corresponding projectors, the Grothendieck
Groups. So in a weak sense we may say that the particular Moduli
space may correspond to those K theoretic objects.

To begin with the construction let us first define the meaning of
graph \cite{R.J.Wilson1979}.

A directed graph G consists of \\
(a) A set V whose elements are termed as vertices.\\
(b) A set E of ordered pairs (u, v) of vertices called edges.\\
V(E) $\equiv$ set of vertices ; \hspace{1cm} E(G) $\equiv$ set of
edges in G.

Now the basic query regarding this point of view which will be a
matter of concern at this stage is that the connection of space
of graphs G with the space of paths on which our principle of
holonomy is based.

To be definitive a directed path P in G is an alternating
sequence of vertices and directed edges say\\
$$P = (v_{0}, e_{1}, v_{1}, e_{2}, v_{2},\ldots e_{n}, v_{n})$$
where $e_{i} = (v_{i-1}, v_{i}) , |P| = n  = $ no of edges.A
Simple path is a path with distinct vertices. A loop has the same
first and last vertices. As a result of these definitions it is
more or less obvious that the space of paths do form a subspace
of the space of graphs G. So though we will be working on the set
of paths. We assume without farthur verification that the results
for the space of graphs do carry over there. The next most
important observation which comes to our hand is that regarding
the construction of graph algebras. A directed graph E consists
of a countable edge set $E^{1}$ and range and source maps $r,s :
E^{1}\rightarrow  E^{0}$. When each vertex receives at most
finetly edges, the graph algebra C$^*$(E) is the universal C$^*$
algebra generated by the mutually orthogonal projections $\{
P_{\theta}: \theta \in E^{0}\}$, partial isometries $\{ s_{e} : e
\in E^{1}\}$ satisfying $s_{e}^{\star}s_{e} = p_{s_{e}}, \forall
e \in E^{1}$ and $ p_{\theta} = \sum s_{e}{\star}s_{e} $ when $
r^{-1}(v) $ is nonempty.

So we arrive at the following definition\\
\textbf{Defn:} Let $ ( \Gamma,\Gamma^{0}, d) $ be a finitely
aligned K graph. A Cuntz Kreiger $ \Gamma $-family
\cite{W.Krieger1980,I.Raeburn1998} is a collection $ \lbrace
t_{\lambda}: \lambda\in \Lambda\rbrace $ of partial isometries
and $\lbrace
 t_{\theta}: \theta\in
\Lambda^{0}\rbrace $ orthogonal projections satisfying \\
\begin{eqnarray}
t_{\lambda}t_{\mu} &=& t_{\lambda\mu}\,when \,s(\lambda) =
\gamma(\mu)\\ \nonumber
 t_{\lambda}^{\ast}t_{\mu} &=&
\sum_{(\alpha\beta)\in\Lambda^{min}(\lambda\mu)}t_{\alpha}t_{\beta}^{\ast}\,
\forall\lambda,\,\mu\,\in\Lambda\\ \nonumber
 t_{\theta}(z) &=& e^{2\pi i \theta}z
 \end{eqnarray}
As a consequence we see that these definitions make C(E) into an algebra by the consequence of the following theorem.\\
\textbf{Theorem:}\\
Let $(\Lambda, \Lambda^{0}, d)$ be a finitely aligned K-graph
which satisfies the following criterion: for each $\theta \in
\Lambda^{0}  \, \exists \,z\in \theta\Lambda^{\leq \infty}\,$
such that $\,\lambda,\mu \in \Lambda\theta \, and \, \lambda
\neq\mu \Rightarrow \lambda z \neq \mu z$\\
Suppose $ \pi $ is a homomorphism of $C^{*}(\Lambda)$ such that $ \pi(s_{k}) \neq 0 \, \forall \theta
\in \Lambda^{0} $
then $ \pi $ is injective.

Now as we have stated the path or in particular the loop space do form a subspace of the graph space and they by dint of
 previous arguments do form a $C^{*}$ algebra.We dont review here
 the basics of C* algebra. Some good references are \cite{Cuntz.J1977,G.J.Murphy1990}

Equipped with all these tools we are now ready to explore into the
moduli space of monopoles as we have promised. The analysis of
section {\ref{secsw}} gives us the required clue as regards the
construction. As we already know that loops give rise to a
representation in terms of gauge fields which give rise to
soliton solutions.
\begin{equation}\label{uni}
 U( \varphi + 2\pi) = h^{-1}(\alpha)
U(\varphi)U(2\pi) h(\alpha)
\end{equation}
 We consider a class of operators in
correspondence to the partial isometries of the $ C^{*}(E) $which
act on the fields
of the moduli space given by the following rule\\
$$ T^{(a)}_{t_{\lambda}}\phi_{a} = e^{-i\Delta t_{\lambda}}
\phi_{t_{\lambda}(a)}$$ In physical terms these operators in some
sense act as if parallel tranporting the field along a loop based
at a point "a". Here $\Delta$  is a parameter.

Now we construct the space of operators $$\mathcal{A} =
\{T^{(a)}_{t_{\lambda}}|  T^{(a)}_{t_{\lambda}}:
\mathrm{M}\rightarrow \mathrm{M} ; \lambda \in \Lambda\}$$
We define\\
 $$T^{(a)}_{t_{\lambda}}(c_{1}\phi_{a} + c_{2}\psi_{a}) = (c_{1}\phi +
 c_{2}\psi)_{t_{\lambda}(a)} = c_{1}\phi_{t_{\lambda}(a)} +
 c_{2}\psi_{t_{\lambda}(a)}$$
 which make this into a vector space.

 Now to make $\mathcal{A}$ into an algebra we properly define the
 product of operators \\
$$ T^{(a)}_{s_{\lambda}}T^{(a)}_{s_{\mu}}\phi_{a} = e^{-i\Delta
[s_{\lambda}, s_{\mu}]} \phi_{s_{\lambda}s_{\mu}(a)} =
e^{-i\Delta [s_{\lambda}, s_{\mu}]} \phi_{s_{\lambda\mu}(a)}$$ We
define the norm on $\mathcal{A}$ by $ \parallel \cdot \parallel :
\mathcal{A} \rightarrow R^{+}$
 $$\parallel
T_{s_{\lambda}}\parallel = sup_{\lambda\in \Lambda}\mid
s_{\lambda}\mid$$ To make $\mathcal{A}$ into a $^*$ algebra let's
define
 $$* : \mathcal{A} \rightarrow \mathcal{A} \,\,\,\, ;
a \rightarrow  a^{*}$$ by $$T_{s_{\lambda}} \rightarrow
T_{s_{\lambda^{*}}}$$ which gives $$ T_{s_{\lambda^{*}}}\phi_{a}
= e^{-i\Delta s_{\lambda^{*}} }\phi(s_{\lambda^{*}}a)$$ with the
above definitions it is easy to show that it becomes a C* algebra
which we explicitly show below.
\begin{eqnarray}\label{alg}
{T_{s_{\lambda^{*}}}}^{*}\phi_{a} &=& [ {e^{-i\Delta
s_{\lambda^{*}}}} \phi(s_{\lambda^{*}}a)]^{*} \nonumber\\
 &=&  e^{-i\Delta s_{\lambda}
}\phi(s_{\lambda}a) \quad[s_{\lambda^{*}} = s_{\lambda}]
\nonumber\\
 &=& T_{s_{\lambda}}\phi_{a} \Rightarrow
T_{s_{\lambda^{*}}}^{*} = T_{s_{\lambda}}
\end{eqnarray}
\begin{eqnarray}\label{}
T_{s_{\lambda^{*}}} \cdot T_{s_{\mu}} &=& {e^{-i\Delta}
[s_{\lambda}, s_{\mu}]}\phi_{ s_{\lambda\mu}}a \nonumber\\
\Rightarrow (T_{s_{\lambda}}\cdot T_{s_{\mu}})^{*} \phi_{a} &=&
{e^{-i\Delta} [s_{\lambda}, s_{\mu}]^{*}}\phi_{
s_{{\lambda}^{*}\mu^{*}}}a \nonumber\\
&=& T_{s_{\mu^{*}}} T_{s_{\lambda^{*}}}\phi_{a} \Rightarrow
(T_{s_{\lambda^{*}}} \cdot T_{s_{\mu}})^{*} =
T_{s_{\mu^{*}}}T_{s_{\lambda^{*}}}\\ \nonumber
\end{eqnarray}
Again
 $$\parallel(T_{s_{\lambda}})^{*}T_{s_{\lambda}}\parallel =
{\parallel T_{s_{\lambda}}\parallel}^{2}$$
\textbf{Theorem:}\\
Let S be a unilateral shift on the Hilbert space $l_{2}(N)$. The
Toeplitz algebra $\mathcal{T}$ is the unital separable C*
subalgebra of $B(l_{2})$ generated by S. The Toeplitz algebra
$\mathcal{T}$ is isomorphic to the C* algebra generated by
$\{{\mathcal T}_{\phi}\mid \phi \in C(\mathcal T)\}$

\textbf{Corollary:}\\
The Algebra $\mathcal{A} = \{T_{s_{\lambda}}\mid T_{s_{\lambda}}:
\mathcal{M} \rightarrow \mathcal{M}\}$ is isomorphic to a
corresponding Toeplitz algebra.\\
\textit{Proof:}  The proof is obvious from the previous
construction and the theorem.

Now we consider the operators  corresponding to the orthogonal
projections $s_{\theta}$ by defining, $$
T_{s_{\theta}}\tilde{\phi_{a}} = s_{\lambda}(e^{2\pi i
\theta}\tilde\phi_{a})$$ So as regards our physical
interpretations the algebra corresponding to the orthogonal
projectors do form a C* algebra induced by the corresponding loop
algebra.

As a consequence of the above propositions the algebra of
operators do decompose into a direct sum  $$ \mathcal{A} =
\mathcal{A_{\infty}} \oplus \mathcal{A_{\varepsilon}}$$ where
$$\mathcal{A_{\infty}} = \{ T_{s_{\lambda}} \mid T_{s_{\lambda}}:
\mathcal{M} \rightarrow \mathcal{M}\quad  T_{s_{\lambda}}
\phi_{a} = e^{i \Delta s_{\lambda}}\phi_{s_{\lambda}}a\}$$
$$\mathcal{A_{\varepsilon}} =  \{ T_{s_{\theta}} \mid
T_{s_{\theta}}: \mathcal{M} \rightarrow \mathcal{M}\quad
T_{s_{\theta}} \phi_{a} = s_{\lambda}(e^{2 \pi i
\theta}\phi_{a})\}$$ where as we have earlier seen that $
\mathcal{A_{\infty}}$ is isomorphic to a suitable Toeplitz
algebra and $\mathcal{A_{\varepsilon}}$ is a rotation algebra
which is a non AF algebra generated by $ '\theta'$ where $0 <
\theta < 1$.

To make contact with the corresponding physics we try for an
interpretation. As it can be understood we try for an algebraic
formalism for the moduli space geometry by taking into account
the loop space of the moduli space of the monopoles and make a
suitable choice of operators as has been motivated by by the
example illucidated by Alice String behaviour. The loops
encircling the singularities of the moduli space can be tracked
by the partial isometries and do correspond to the algebra
$\mathcal{A_{\infty}}$ and the loops which do not encircle it do
give rise to $\mathcal{A_{\varepsilon}}$

As we have constructed an algebraic decomposition of the monopole
moduli, our next objective is the construction of some invariants
related to the loop space algebra. So the basic goal to
accomplish is schematically shown below  $$ \mathcal{M}
\Longrightarrow \mathcal{L} \Longrightarrow C (\mathcal{L})
\Longrightarrow Z_{0}$$

where $ \mathcal{M}$ denotes the monopole space with which we start with,
$\mathcal{L}$ the corresponding loop space, $C(\mathcal{L})$ as has been noted previously the C*
algebra of the loop space,
 $Z_{0}$ the cherished invariants or groups which we aim to construct. So effectively we wish to
  associate a group
 corresponding to the given moduli space. This construction reminds us of the basic tenants of
  sheaf theory in algebraic
 geometry where we associate an abelian group corresponding to an open set. So as wee see the
  decomposition of the loop
 space has resulted into two parts one consisting of the Toeplitz algebra and the other to a
 rotation algebra.
 Now we will concentrate on a general mechanism of how to generate invariants of the algebra
 \cite{M.F.Atiyah1967,Karoubi.M1978}
  ,extension to our case
 will be considered thereafter.
\subsection{Generating Invariants of the algebra}
Now as we constructed the C* algebra corresponding to the loop
space of the moduli space which has been constructed using graphs
it is an important question at this stage to ask whether there
exist any any invariants associated with this algebra. In general
to each C* algebra $\mathcal A$ whether unital or non unital there
exist two K groups \cite{J.Cuntz1982}which are $\mathcal
K_{0}(\mathcal A)$ and $\mathcal K_{1}(\mathcal A)$. They are
constructed using matrices over $\mathcal A$. In case of
$\mathcal K_{0}$ the matrices should be projections and for
$\mathcal K_{1}$ they are unitaries. In this section at first we
will consider the general scheme and results for computing the
relevant groups from an underlying C* algebra and then comment on
the possible K groups
and their properties in our case of moduli space.\\
\textbf{Theorem:} \\
Let $ \mathcal{A_{\infty}}$ be the Toeplitz algebra (generated by
the unilateral shift operator S). Let  $\mathcal{K}$ be the
 set of compact operators on a separable infinite dimensional Hilbert space $H =
 l_{2}$ Then,\\
i)$ \mathcal{K}$ is an ideal in $
\mathcal{A_{\infty}}$\hspace{0.1cm} ii) There exists an exact
sequence $$ 0 \Longrightarrow \mathcal{K}\Longrightarrow
\mathcal{A_{\infty}} \Longrightarrow C(\mathcal{A_{\infty}})$$
\textbf{Proof:}\\
i) If we compose the quotient map $\mathcal A_{\infty} \rightarrow
\mathcal A_{\infty}/\mathcal{K} \simeq C(\mathcal A_{\infty})$
with evaluation in $1\in \mathcal A_{\infty}$ we get a morphism $
f: \mathcal A_{\infty} \rightarrow C$ whose kernel
$\widetilde{\mathcal{A}}$ is the C* algebra generated by $1-S$,
where S is the unilateral shift operator as an ideal and $
\widetilde{\mathcal A}/\mathcal K \cong C_{0}(R)$ \\ii) If $M_{n}$
are modules over a unital ring R and $\mu_{n}: M_{n}\rightarrow
M_{n-1}$ are module maps then the sequence $$ \cdots
\stackrel{\mu_{n-2}}\Longrightarrow M_{n-1}
\stackrel{\mu_{n-1}}\Longrightarrow M_{n}
\stackrel{\mu_{n}}\Longrightarrow M_{n+1}
\stackrel{\mu_{n+1}}\Longrightarrow \cdots$$ is called exact when
for each pair of consequetive maps the image of the first is the
kernel of the second i.e, ker  $\mu_{n-1}$ = Range  $\mu_{n}$\\
The standard exact sequence of C* algebras occur when A is an
ideal in B. If $ \pi : B \rightarrow B/A $ is the projection onto
the quotient and $ i : A \hookrightarrow B $ is the inclusion map
then $$ 0 \rightarrow A \stackrel{i}\hookrightarrow B
\stackrel{A}\rightarrow B/A \rightarrow 0 $$ is exact.\\
Taking the result of part (i) we see that there is an exact
sequence of $$ 0 \rightarrow \mathcal K
\stackrel{i}\hookrightarrow \mathcal
A_{1}\stackrel{\pi}\rightarrow C(\mathcal A_{1}) \rightarrow 0$$
\textbf{Lemma:}Let J be an ideal in $ \mathcal A_{1}$. If $x\in
\mathcal A_{1}$, then $$ x\in J \Leftrightarrow \pi_{J}(x) = 0
\quad\quad; x\in J^{+}\Leftrightarrow \pi_{J}(x) = C$$
\textbf{Proof:} If $x = a + \lambda$ $a\in \mathcal A_{1}$,
$\lambda \in C$ then $ \pi_{J}(x) = \pi_{J}(a) + \lambda$ by the
definition of ideal in $ \mathcal A_{1}$.\vspace{.5cm}
\textbf{Theorem:} Half Exactness of $K_{0}$\\
 An exact sequence
$0 \rightarrow \mathcal K \stackrel{i}\hookrightarrow \mathcal
A_{1}\stackrel{\pi}\rightarrow \mathcal A_{1}/{\mathcal{K}}
\rightarrow 0$, where $\mathcal{K}$ is an ideal in $\mathcal{A}$
induces a short sequence of $ \mathcal{K_{0}}$ groups. $$\mathcal
K _{0}{\mathcal K}\stackrel{i_{*}}\hookrightarrow \mathcal
K_{0}(\mathcal A_{1})\stackrel{\pi_{*}}\rightarrow \mathcal K_{0}(
\mathcal A_{1}/{\mathcal{K}})$$ \textbf{Proof:} We here Prove that
Let $x\in \mathcal K_{0}(J)$,so x can be decomposed as $ x = [p] -
[p_{n}]$ where p is the projection ,$p\in
\mathcal{M_{\infty}}(J^{+})$ So $$ \pi_{*} \circ i_{*}(x) =
[\pi_{J}(p)] - [\pi_{J}(p_{n})] = [p_{n}] - [p_{n}] = 0$$ Hence
Range  $i_{*} \subset$ Ker  $\pi_{*}$ Along somewhat similar lines
we can prove the converse namely Ker  $\pi_{*} \subset$ Range
$i_{*}$ to prove the theorem.\\
 The following theorem is the counterpart of the previous Theorem
 in the $\mathcal K_{0}$ case. which we state without Proof.\\
\textbf{Theorem:} Half Exactness of $\mathcal K_{1}$\\
The exact sequence $0 \rightarrow \mathcal J
\stackrel{i}\hookrightarrow \mathcal A\stackrel{\pi}\rightarrow
\mathcal A/{\mathcal{J}} \rightarrow 0$ where $\mathcal{J}$ is an
ideal in $\mathcal{A}$ induces a short sequence of $
\mathcal{K_{1}}$ groups: $$\mathcal K _{1}{\mathcal
J}\stackrel{i_{*}}\hookrightarrow \mathcal K_{1}(\mathcal
A)\stackrel{\pi_{*}}\rightarrow \mathcal K_{1}( \mathcal
A/{\mathcal{J}})$$ Now we state the following theorems without
proof which will be essential to construct the K groups for the
moduli space.\\
\textbf{Theorem:}\\
Let $\mathcal J$ be an ideal in $\mathcal A$ . The following
sequence is exact everywhere
 $$\mathcal K _{1}(\mathcal
J)\stackrel{i_{*}}\hookrightarrow \mathcal K_{1}(\mathcal
A)\stackrel{\pi_{*}}\rightarrow \mathcal K_{1}( \mathcal
A/{\mathcal{J}})\stackrel{\delta}\rightarrow \mathcal K
_{0}(\mathcal J)\stackrel{i_{*}}\hookrightarrow \mathcal
K_{0}(\mathcal A)\stackrel{\pi_{*}}\rightarrow \mathcal K_{0}(
\mathcal A/{\mathcal{J}})$$ where $\delta$ is the index map
$\delta:\mathcal K_{1}( \mathcal A/{\mathcal{J}})\rightarrow
\mathcal K _{0}(\mathcal J) $defined by $\delta (x) =
[\varepsilon p_{n}] - [p_{n}]$ where $\varepsilon$ is a unitary
lift of the C* algebra A.\\
\textbf{Theorem:}\\
The index map $\delta$ can be shown to be an isomorphism and it
implies that $$\mathcal K _{1}(\mathcal A_{1}) = 0 ; \quad\quad
\mathcal K _{0}(\mathcal A_{1}) = \mathcal Z$$ With the above
results we are in a position to deal with the K groups
corresponding to the loop space algebra of the moduli space. But
before that we will require some standard result from PV
sequences which will be helpful in understanding the structure of
the Grothendieck groups to be considered later. In the discussions
below we consider C* algebra with unit. Computation of K groups
from a given as it can be well understood is a highly nontrivial
job and it naturally depends on the respective algebra concerned.
At present there are standard prescriptions which can be used to
find out K theory of C* algebras which involve crossed products.
PV sequence is related with a correspondence of a unital C*
algebra A and every automorphism $\alpha$ of A a cyclic six term
exact sequence, which can be expressed in the following
commutative diagram.
 $$
\xymatrix{\mathcal K_{1}(A)\ar[r]^{i - \alpha_{*}}& \mathcal
K_{1}(A)\ar[r]^{i_{*}}& \mathcal K_{1}(A \times_{\alpha} \mathcal
U)\ar[d]\\
\mathcal K_{0}(A \times_{\alpha} \mathcal U)\ar[u] & \mathcal
K_{0}(A)\ar[l]^{i_{*}} & \mathcal K_{0}(A)\ar[l]^{i - \alpha_{*}}
} $$ where $i: \mathcal A \hookrightarrow \mathcal A
\times_{\alpha} \mathcal U$ is the natural embedding, $\mathcal U$
is a suitable algebra for the required purpose. \\
Now to return to our case, motivated by the Alice string behaviour
we are interested in the unitary operator implementation of
(\ref{uni}) the corresponding C* algebra of the loop space
reflects the underlying property of the gauge fields which has
been previously argued to be decomposable into the sum of
Toeplitz and Rotation algebras. But it should be clear that in
general this may not be at all the case. Depending on variety of
physical situations the structure of gauge fields residing in the
moduli space can have a wide variety of manifestations, for
example D-brane moduli space in TypeII String Theories
accomplishes a different
character\cite{G.Moore1996,Moore1998,S.Stanciu2001}. In our case
we have been lucky enough utilising PV sequences to show that the
moduli space
gives rise to some K theoretic groups. \\
Let us consider an automorphism $ \mathcal P_{\lambda}$ of the
algebra of continuos functions on a unit circle I as $\mathcal C(I)$\\
So, for the decomposition of $\mathcal{A_{\varepsilon}}$ we define
$$ \mathcal P_{\lambda}(T_{s_{\theta}})(x) =
T_{s_{\theta}}(e^{2\pi i\lambda}x)\quad ;\quad\quad
T\in\mathcal{A_{\varepsilon}},\quad x\in I$$ But we know that $$
T_{s_{\theta}}\tilde{\phi_{a}} = s_{\lambda}(e^{2\pi i
\theta}\tilde\phi_{a})$$ Thereby combining the above two
equations we see that $\mathcal{A_{\varepsilon}}$ can be realized
as the crossed product $\mathcal C(I) \times_{ \mathcal
P_{\lambda}} \mathcal S$ where $\mathcal S^{1}$ is the unit disc.
The PV sequence becomes
 $$
\xymatrix{\mathcal S^{1}\ar[r]^{0}& \mathcal S^{1}\ar[r]&
\mathcal K_{0} \mathcal \mathcal{A_{\varepsilon}}
\ar[d]\\
\mathcal K_{1}\mathcal{A_{\varepsilon}}\ar[u] & \mathcal
S^{1}\ar[l]^{0} & \mathcal S^{1}\ar[l]} $$ So the result follows
directly as $\mathcal K_{0}\mathcal{A_{\varepsilon}}\simeq
\mathcal K_{1}\mathcal{A_{\varepsilon}} \simeq \mathcal S^{1}
\bigoplus \mathcal S^{1}$ \\
In the second case after identifying the algebra to be one of
Toeplitz algebra the analysis is much more straightforward and
well known.\\
As we have clarified earlier that $\mathcal{A_{\infty}}$ is a
Toeplitz algebra It can be verified that $\mathcal{A_{\infty}}$
contains $\mathcal K$ as an ideal since each of the matrix
elements in ${\mathcal M}_{n} \subseteq {\mathcal B}{l_{n}}$ may
be expressed as a polynomial of $S$ and $S*$ where $S*$ is the
adjoint of $S$. From this it can be proved that the following
sequence is exact $$0\rightarrow \mathcal K
\stackrel{i}\hookrightarrow \mathcal{A_{\infty}}
\stackrel{\pi}\rightarrow C(\mathcal{A_{\infty}})\rightarrow 0$$
To compute the K groups corresponding to the associated Toeplitz
algebra we state the following theorem without proof \\
\textbf{Theorem:}\\
Each Exact Sequence $$0\rightarrow  X \rightarrow Y \rightarrow
Z\rightarrow 0$$ induces a cyclic six term exact sequence
 $$
\xymatrix{\mathcal K(X)\ar[r]& \mathcal K(Y)\ar[r]& \mathcal K(Z)
\ar[d]\\
\mathcal K(SZ) \ar[u] & \mathcal K(SY)\ar[l] & \mathcal
K(SX)\ar[l]} $$ where $S$  is the unilateral shift operator. it
will be worthwhile to make some comments here. It is known that
there is an isomorphism $\mathcal K \simeq  \mathcal K S^{2}$ .
The above theorem then can be readily replaced with $\mathcal K,
\mathcal{A_{\infty}}, C(\mathcal{A_{\infty}})$ for $X, Y, Z$

Since we have verified that $\mathcal K$ is an ideal to
$\mathcal{A_{\infty}}$ we can define an index map $$\delta :
\mathcal K_{1}(\mathcal{A_{\infty}}/\mathcal K) \rightarrow
\mathcal K_{0}(\mathcal K)$$ It may be verified  that $\delta$ is
an isomorphism and thereby $$\mathcal
K_{0}(\mathcal{A_{\infty}})\approx \mathcal Z\quad;
\quad\quad\quad \mathcal K_{1}(\mathcal{A_{\infty}})
 = 0$$
Thus in accordance with our studies in (\ref{geomod}) we find
that in general that the loop space of the monopole moduli  will
give rise to two group invariants given by above. Though it
should be clear from the above that the construction has been
motivated by Eq(\ref{alice}) and the important connection between
twisted Alice loops and monopoles. This construction does not
imply that given an arbitrary moduli space it will be at all
possible to find the K-theoretic invariants regarding that space.
It is at all clear whether such a construction is at all possible
or not.
\section*{Acknowledgements}
The author wishes to acknowledge the hospitality of ICTP where some portions of the present work was conceived during his visit there during July 2002.Some friutful discussions with Prof A.O.Kuku is also gratefully acknowledged.
\newpage


\end{document}